\documentclass[aps,letterpaper,twocolumn,prl,footinbib,floatfix]{revtex4}
\usepackage{amsmath}
\usepackage{amsfonts}
\usepackage{amssymb}
\usepackage[scanall]{psfrag}
\usepackage{psfrag}
\usepackage{graphicx}
\usepackage{color}

\begin{document}
\newcommand{\of}[1]{\left( #1 \right)}
\newcommand{\sqof}[1]{\left[ #1 \right]}
\newcommand{\abs}[1]{\left| #1 \right|}
\newcommand{\avg}[1]{\left< #1 \right>}
\newcommand{\cuof}[1]{\left \{ #1 \right \} }
\newcommand{\bra}[1]{\left < #1 \right | }
\newcommand{\ket}[1]{\left | #1 \right > }
\newcommand{\pil}{\frac{\pi}{L}}
\newcommand{\bx}{\mathbf{x}}
\newcommand{\by}{\mathbf{y}}
\newcommand{\bk}{\mathbf{k}}
\newcommand{\bp}{\mathbf{p}}
\newcommand{\bl}{\mathbf{l}}
\newcommand{\bq}{\mathbf{q}}
\newcommand{\bs}{\mathbf{s}}
\newcommand{\psibar}{\overline{\psi}}
\newcommand{\svec}{\overrightarrow{\sigma}}
\newcommand{\dvec}{\overrightarrow{\partial}}
\newcommand{\bA}{\mathbf{A}}
\newcommand{\bdelta}{\mathbf{\delta}}
\newcommand{\bK}{\mathbf{K}}
\newcommand{\bQ}{\mathbf{Q}}
\newcommand{\bG}{\mathbf{G}}
\newcommand{\bw}{\mathbf{w}}
\newcommand{\bL}{\mathbf{L}}
\newcommand{\ohat}{\widehat{O}}
\newcommand{\up}{\uparrow}
\newcommand{\down}{\downarrow}
\newcommand{\MM}{\mathcal{M}}
\newcommand{\tX}{\tilde{X}}
\newcommand{\tY}{\tilde{Y}}
\newcommand{\tZ}{\tilde{Z}}
\author{Eliot Kapit}
\affiliation{Department of Physics and Engineering Physics, Tulane University, New Orleans, LA 70118}

\title{Error-transparent quantum gates for small logical qubit architectures }

\begin{abstract}

One of the largest obstacles to building a quantum computer is gate error, where the physical evolution of the state of a qubit or group of qubits during a gate operation does not match the intended unitary transformation. Gate error stems from a combination of control errors and random single qubit errors from interaction with the environment. While great strides have been made in mitigating control errors, intrinsic qubit error remains a serious problem that sets the primary limit for gate fidelity in modern superconducting qubit architectures. Simultaneously, recent developments of small error-corrected logical qubit devices promise significant increases in logical state lifetime, but translating those improvements into increases in gate fidelity is a complex challenge. In this Letter, we propose a new formalism for implementing gates on and between small logical qubit devices which inherit the parent device's tolerance to single qubit errors which occur at any time before or during the gate. Using a standard phenomenological noise model for superconducting qubits, we demonstrate a universal one- and two-qubit gate set with error rates an order of magnitude lower than those for equivalent operations on single qubits or pairs of qubits, running for the same total duration. The effective logical gate error rate in these models displays superlinear error reduction with linear increases in single qubit lifetime, proving that passive error correction is capable of increasing gate fidelity. These developments further suggest that incorporating small logical qubits into a measurement based code could substantially improve code performance.

\end{abstract}

\maketitle

To build a fault-tolerant, error-corrected quantum computer, every operation in the code (one- and two-qubit gates, state preparation, measurement and idling) must be performed to extremely high fidelity \cite{fowlersurface,terhal2015}. While the requisite fidelities have been achieved in single qubit gates \cite{chowgambetta2012,barendskelly2014}, improving two-qubit gate performance is considerably more difficult \cite{kellybarends2014,mckayfilipp2016,paikmezzacapo2016,blumoffchou2016}, with experimentally realized gate error not far below the threshold rate. Further, the classical processing required for a code involving tens or hundreds of millions of physical qubits is daunting, and increasing the cycle time to reduce this burden increases error rates, further degrading code performance. An improved qubit primitive capable of much higher two-qubit gate fidelity and state lifetimes could thus make it dramatically easier to implement a topological code.

We propose to address both challenges by extending our recent proposal for a small, passively error corrected quantum device \cite{kapit2016}, called the ``Very Small Logical Qubit." Small logical qubit circuits have attracted much interest in recent years \cite{mirrahimileghtas2014,sunpetrenko2014,leghtastouzard2015,cohenmirrahimi2014}, including the first experimental demonstration of a quantum error correction protocol that exceeds breakeven \cite{ofekpetrenko2016}. However, due to the larger and more complex Hilbert space, error correction that increases idle lifetime does not necessarily improve gate fidelity. To achieve this, we propose a new formalism for engineering ``error-transparent" quantum gates \footnote{We refer to these gates as ``error-transparent," since they are resilient to single qubit error which occurs before or during their operation. Error transparency is distinct from the more general notion of fault tolerance, as fault tolerance is typically interpreted as the ability to exponentially reduce logical error rates from a polynomial increase in circuit complexity. The small logical qubit circuits considered here do not have an obvious scaling path, but perform extremely well against single error events, and could potentially improve code performance by replacing single qubits in a larger measurement-based code.}, where the physical Hamiltonian implementing the gate is carefully tailored such that it commutes with single-qubit errors (in this case, photon losses) when acting on the logical state manifold, \textit{at all times during the gate operation}. This criteria ensures that subsequent error correction will recover the correct (transformed) logical state regardless of when the error occurred during the gate operation. The error rate of such a gate in the ideal limit would thus decrease as $T_g T_R/ T_1^2$ (where $T_g, T_R$ and $T_1$ are the gate, error correction and random error timescales, respectively), leading to large improvements in gate fidelity, as random single qubit errors are the current limiting factor in well-designed architectures \cite{kellybarends2014}. These developments are similar to recent work for cat codes, where robust gate \cite{mirrahimileghtas2014,albertshu2016} and measurement \cite{cohensmith2016} protocols have been proposed, though these schemes fall short of a complete universal gate set which is insensitive to single errors that occur at any random point during any gate. We will describe how to implement an error-transparent gate set for the VSLQ architecture, and benchmark its performance through numerical simulation. We will demonstrate super-linear decreases in gate error with increased $T_1$ and show that two-qubit gate error rates in the low $10^{-4}$ range are achievable without further increases in base qubit coherence.  

\textbf{The Very Small Logical Qubit:} We shall blueprint a realistic implementation of an error-transparent gate set in the VSLQ architecture \cite{kapit2016}. A VSLQ consists of a pair of transmons coupled by a tunable, flux driven coupler \cite{kapit2015} driven at high frequencies to coherently drive two- and four-photon transitions. Defining $\tX_{i} \equiv \of{a_i^\dagger a_i^\dagger + a_i a_i}/\sqrt{2}$ and $P_{i}^{j}$ to be the projector onto states where object $i$ contains exactly $j$ photons, the rotating frame VSLQ Hamiltonian, in the three level basis \footnote{The continuously increasing nonlinearity allows us to effectively ignore the $\ket{3}$ and $\ket{4}$ states of each transmon. Mixing with these states will create small higher order corrections, which can be easily compensated by additional signals to counter their effects. For clarity and simplicity we will ignore these effects here.} of the left and right qubits $l$ and $r$ is given by:
\begin{eqnarray}\label{HP}
H_P = - W \tX_l \tX_r + \frac{\delta}{2} \of{ P_l^1 + P_r^1}
\end{eqnarray}
The ground states of the VSLQ are the two states satisfying $\tX_l \tX_r = 1$. For the simulations in this paper, we used $W = 25$ MHz and $\delta = 300$ MHz (both $\times 2 \pi$). Given the phenomenological noise model for superconducting qubits of low-frequency phase noise \cite{martinisnam2003,ithiercollin2005,yoshiharaharrabi2006,bylandergustavsson2011,antonmuller2012,yangustavsson2013,paladinogalperin2014,omalleykelly2015} and white noise photon loss, when coupled to additional lossy elements the VSLQ acts as a logical qubit protected against all single qubit error channels. Specifically, we introduce two additional lossy ``shadow" qubits \cite{kapithafezi2014}, with circuit Hamiltonian
\begin{eqnarray}\label{HPS}
H &=& H_P + H_S + H_{PS}, \; H_{S} = \omega_S \of{a_{Sl}^\dagger a_{Sl} + a_{Sr}^\dagger a_{Sr}}  , \nonumber \\
H_{PS} &=& \Omega \of{t} \of{a_l^\dagger a_{Sl}^\dagger + a_r^\dagger a_{Sr}^\dagger + {\rm (H.c.)}  } 
\end{eqnarray} 
By careful tuning of $\omega_S$ and $\Omega \of{t}$, a photon loss in a primary qubit can be converted to an excitation in a shadow qubit, returning the VSLQ to its original logical state. By then introducing a fast loss rate for the shadow qubits, the shadow qubit excitation can be eliminated, returning the system to its rotating frame ground state and completely eliminating the error. While phase errors cannot be corrected through this mechanism, the large $W$ term introduces an energy penalty for phase errors. Since phase noise is low-frequency dominated, it is thus strongly suppressed-- see the Supplemental Material of this paper for quantitative simulations.

\begin{figure}
\includegraphics[width=3.25in]{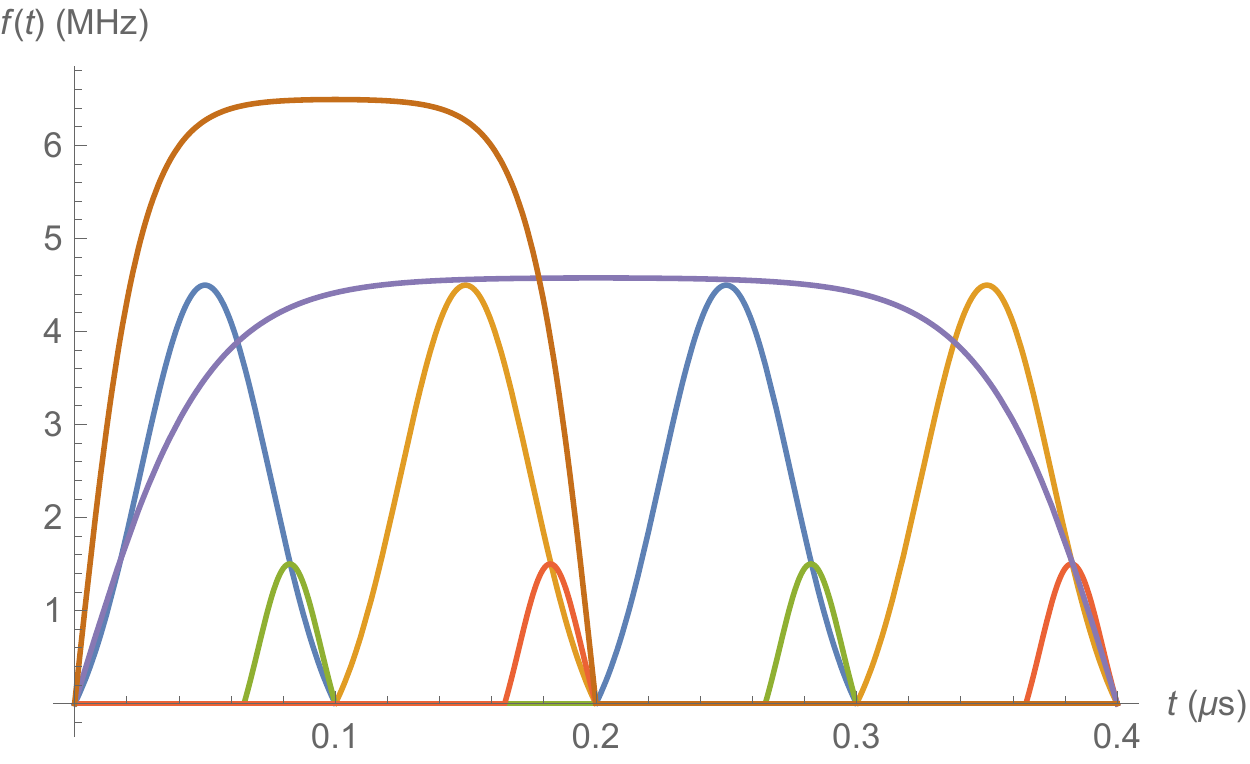}
\caption{(Color online) Gate envelope functions, for two-qubit gates spread out over two or four 100 ns error correction cycles. The blue (orange) curves are the time-dependent Rabi frequencies error correction pulses on the left (right) qubit of each VSLQ copy. Green and red curves plot the $g_{1}$ and $g_{2}$ coefficients in the timed XCX operation (\ref{timedXCX}), which apply $\tX_{lA} \tX_{lB}$ and $\tX_{rA} \tX_{rB}$ terms to couple the two copies. Finally, the purple and brown curves plot the ZZ coefficient $g \of{t}$ in (\ref{fixCZ}) for gate durations of 200 and 400 ns, which enacts an entangling CZZ gate. As this gate is generated by second order terms in perturbation theory, the bare coupling is larger than the XCX terms (which commute with $H_P$); since $W = 25$ MHz the total phase rotation of $\pi/4$ is identical. These waveforms were used in the fidelity simulations of FIG.~\ref{2Qfig}, and are simple Gaussian (EC and XCX) or quadratic (CZZ) profiles which could likely be further improved through numerical optimization.}\label{gatefig}
\end{figure}

\textit{Pulsed error correction and idle error:} In the original VSLQ work, the error correction was continuously applied with constant $\Omega$ and shadow qubit loss rate $\Gamma_S$. However, performance can be improved by running these error correction drives as discrete pulses (FIG.~\ref{gatefig}). In this scheme the shadow qubit lifetime is by default set equal to the primary qubits, but can be rapidly adjusted to a fast loss rate by adjusting its energy to be close to that of a fast readout resonator. The error correction cycle runs as follows: the shadow qubits are set to a low loss rate, the error correction drive is turned on for a full photon loss correction, then the shadow qubit is set to a very fast loss rate with $\Gamma_S$ in the tens of MHz (either though a controllable detuning or driven state transfer to a lossy resonator). This protocol does not dramatically decrease idle error compared to well-chosen continuous drive parameters, but it can have more pronounced effects on gate fidelity, allowing us to implement the timed XCX gate described below. 

\textbf{Error-transparent gates for a single VLSQ:} To generate an error-transparent single qubit gate set for the VSLQ, we need to construct two operators $X_L$ and $Z_L$, which we can apply in combination to produce rotations between any two points on the device's Bloch sphere. A natural pair of operators is given by:
\begin{eqnarray}\label{opbare}
X_L^{\rm{(bare)}} = \tilde{X}_l, \; Z_L^{\rm{(bare)}} = \tilde{Z}_l \tilde{Z}_r; \; \of{\tZ_i \equiv P_i^2 - P_i^0}.
\end{eqnarray}
These operators commute with $H_P$ and anticommute with each other, and sequences of partial rotations constructed from them can implement arbitrary rotations in the logical manifold. 

However, these operators are not error-tolerant, since the bare operators $\tX_i$ and $\tZ_i$ return zero acting on a $\ket{1}$ state, and their commutator with the single photon loss operator $a_i$ thus has $O \of{1}$ matrix elements when acting on the logical state manifold. If a photon loss occurs during a gate, the desired operation will not be continuously applied to the VSLQ until the photon loss is repaired. Since the time between the initial loss and its correction is not measurable in the circuit, an unknown fraction of the gate operation is not applied, producing an unheralded quantum error. To construct error-transparent versions,  modify our operators by defining:
\begin{eqnarray}\label{XZtol}
X_L = \tX_l + P_l^1 \tX_r, \; Z_L = \tZ_l' \tZ_r'; \\
 \of{\tZ_i' \equiv P_i^2 + P_i^1 - P_i^0}. \nonumber
\end{eqnarray}
Both of these operations can be implemented by adding additional signals through the VSLQ's central SQUID, and we shall see that they suffer no loss of fidelity from a single photon loss in either qubit. 

We first consider $X_L$, and imagine that a photon loss occurs in the $r$ qubit during the application of $X_L$ as a gate Hamiltonian. Since there are by default no $\ket{1}$ states in the logical state manifold, the $P_l^1 \tX_r$ term returns zero, and $\sqof{\tX_l,a_r}=0$ trivially, so for $\ket{\psi_L}$ in the logical state manifold $\sqof{a_r,X_L}\ket{\psi_L} = 0$. Similarly, if a photon is lost from the left qubit, the $\tX_l$ term returns zero, but since the logical states are defined by $\tX_l \tX_r = +1$, $\tX_l \ket{\psi_L} = \tX_r \ket{\psi_L}$ and thus the system evolves identically under $P_l^1 \tX_r$, and $\sqof{a_l,X_L}\ket{\psi_L} = 0$ as well. Of course, if two or more photons are lost during the gate operation a logical error will occur, so the gate error should shrink as nearly $T_g T_R/T_{1P}^2$ as $T_{1P}$ grows.

\begin{figure}
\includegraphics[width=3.25in]{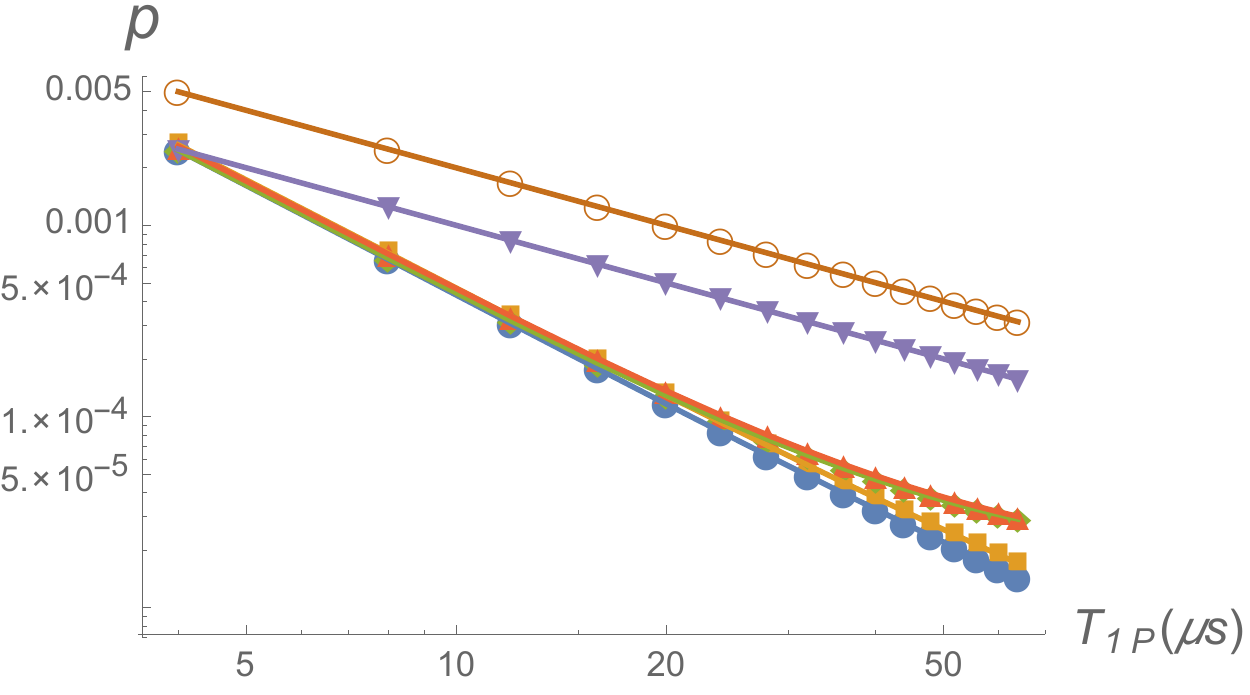}
\caption{(Color online) Fidelity for single-VSLQ error-transparent gate operations using the error-transparent operators (\ref{XZtol}), averaged over the logical Bloch sphere, with pulsed error correction drives and a total EC cycle/gate duration of 200 ns. Here, we plot error rates for idling (blue, filled circles), logical $X$ (gold, filled squares), logical $Z$ (green, diamonds) and logical Hadamard (brown, filled squares; nearly identical to $Z_L$ error rate). For comparison, we include default error rates $1-e^{-T_{g}/2T_{1P}}$ for gate durations of 20 (purple, triangles) and 40 (brown, open circles) ns, assuming no intrinsic gate error. }\label{1Qfid}
\end{figure}

Let us now consider $Z_L$. Since our error model assumes photon losses but no photon addition, if one of the transmons is in a $\ket{1}$ state it decayed from a $\ket{2}$ state in the logical state manifold. As $\tZ'$ returns 1 on both $\ket{1}$ and $\ket{2}$, evolution of a logical state under the operator $\tZ_l ' \tZ_r'$ is unchanged by a single photon loss in either qubit. $Z_L$ is thus similarly protected against single photon losses as $X_L$ is. The performance of these gates against photon loss is shown in FIG.~\ref{1Qfid}. To make the computation tractable, our simulations restricted the VSLQ transmons to the three-level basis and assumed perfect implementation of the error-tolerant operators. The only significant error source in our simulations was thus random photon loss, as control error is negligible for the long gate durations considered. Errors due to the effect of higher levels are very small and can be eliminated by numerical  optimization schemes such as GRAPE \cite{khanejareiss2005}.

\textit{Two qubit gates: Opaque operations timed with error correction:} Implementing a realistic two-VSLQ entangling gate based on the error tolerant operators (\ref{XZtol}) is a subtle challenge. The essential reason for this is that the error tolerant $X_L$ and $Z_L$ are constructed from two-qubit operations, and products of them acting on two VSLQ copies involve three- and four-qubit operations that are difficult to implement. One could engineer these operations using a gadget construction as in \cite{kapitchalker2015}, but doing so increases circuit complexity and the gadget degrees of freedom introduce additional error channels. We will thus avoid this route in this Letter.

Our first method for engineering error-transparent gates is to use the bare, ``opaque" operators $\tX_{lA}$ and $\tX_{lB}$ (for VSLQ copies $A$ and $B$), but timing their operation to coincide with when error correction pulses minimize the instantaneous likelihood of finding a qubit in a $\ket{1}$ state. We let our ideal entangling ``XCX" gate be defined as:
\begin{eqnarray}\label{defXCX}
{\rm XCX} = \exp \sqof{ i \frac{\pi}{4} \of{ X_{LA} - X_{LB} - X_{LA} X_{LB} }}
\end{eqnarray}
The single-qubit parts of (\ref{defXCX}) can be implemented with error-transparent operations, but the entangling two-qubit portion cannot without including four-body terms. So in our physical gate, we apply a pulse 
\begin{eqnarray}\label{timedXCX}
f \of{t} \of{ X_{LA} - X_{LB} } - \of{g_{1} \of{t} \tX_{lA} \tX_{lB} + g_{2} \of{t} \tX_{rA} \tX_{rB}}
\end{eqnarray}
where the $f$ and $g$ terms are gate envelope functions as shown in FIG.~\ref{gatefig}. Note that even in absence of additional terms, these pulses may have an advantage over ordinary two-qubit gates between single transmons, since in those cases, a single photon loss which occurs at any time during the gate is a logical error that ruins the fidelity of the gate. In contrast, a photon loss which occurs during a $\tX$ operation prevents any further state evolution through the $\tX$ operator until the photon loss is corrected, but does not affect the other qubit in the VSLQ and thus does not prevent the parent state from being recovered. In other words, single photon losses cause the gate to be only partially applied, with a fidelity loss that depends on when the photon loss occurs during the gate process. From this we can draw two conclusions: first, that photon losses which occur toward the end of a gate (after most of the gate operation has already occurred) will do little to reduce the fidelity, and second, if error correction is applied during the gate operation it can halt further fidelity loss from a prior photon loss.

\begin{figure}
\includegraphics[width=3.25in]{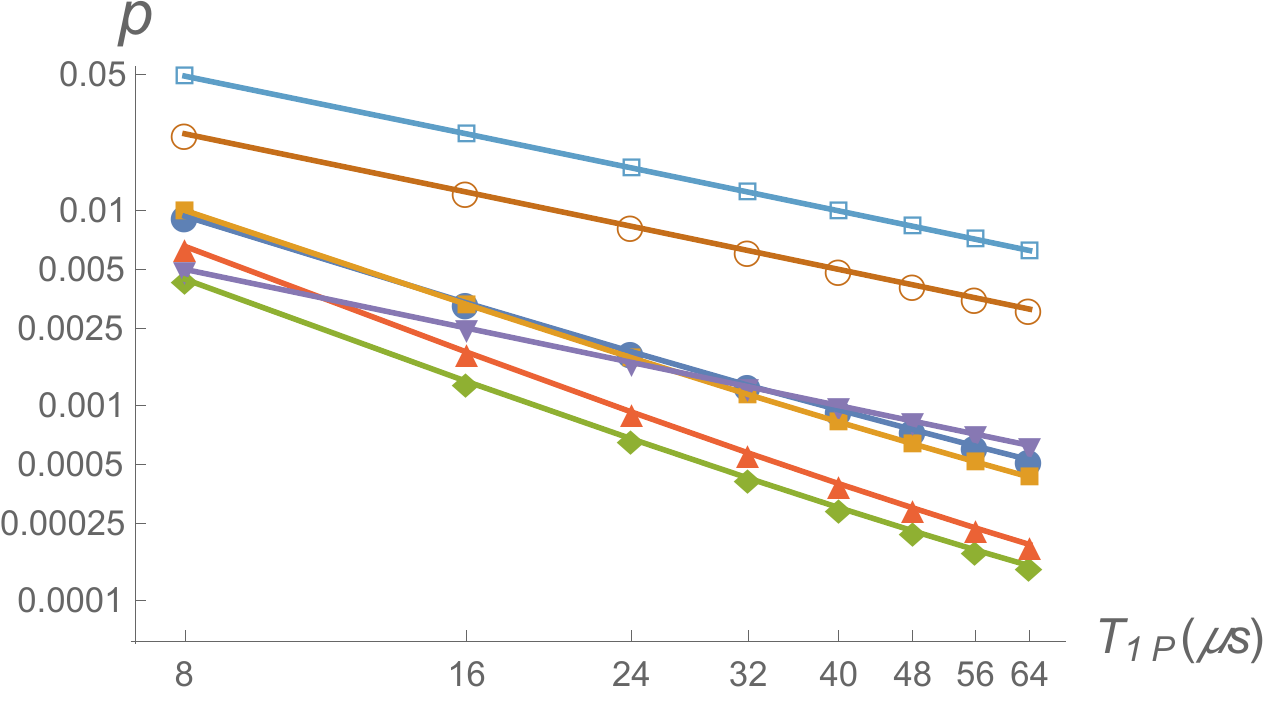}
\caption{(Color online) Fidelity of two-qubit gates, for photon loss rate $T_{1P}$ between 8 and 64 $\mu s$. Here we plot the average error rate $p$ per two-qubit gate for XCX split over two EC pulses (blue, filled circles), XCX split over four EC pulses (orange, squares), CZZ split over two EC pulses (green, diamonds) and four pulses (red, triangles), and bare two-qubit gate error $1-e^{-T_{g}/T_{1P}}$ for $T_{g} =$ 40 ns (purple, triangles), 200 ns (brown, open circles), and 400 ns (blue, open squares). The bare two-qubit gate error is included for comparison purposes and plots the expected gate error from single qubit photon losses occurring during an ordinary two-qubit gate such as CZ, with dephasing and control errors absent. CZZ gates with continuous error correction (not shown) have slightly worse scaling; their error rate is nearly identical at $T_{1P}= 8 {\rm \mu s}$ but is about 60 \% higher at $T_{1P} = 64 {\rm \mu s}$.}\label{2Qfig}
\end{figure}

Our numerical simulations support this prediction. For Gaussian gate and EC pulses \footnote{Due to the computational cost of simulating each gate operation, we did not attempt sophisticated optimization protocols such as GRAPE, and instead chose our gate waveforms based on heuristics and simulations of single-VSLQ gates. Undoubtedly, using such protocols could improve the performance of the gate beyond what we show here.} the best timing we were able to achieve was to apply the $\tX_{lA} \tX_{lB}$ pulses in approximately last third of each EC pulse, with the total gate operation spread out over multiple EC pulses. The results of our simulations are shown in FIG.~\ref{2Qfig}, showing superlinear reductions in gate error with linear increases in $T_{1P}$, with a net error rate of $p \simeq 5.3 \times 10^{-4}$ for $T_{1P} = 64 {\rm \mu s}$. Gate fidelity was found by evolving the system's Lindblad equation \cite{gardinerzoller} until the decay rate equilibrates (eliminating spurious short-time behavior), performing the gate and averaging the resulting error rate over all thirty-six combinations of initial $X,Y,Z$ eigenstates the two VSLQ copies; for further details, see the Supplemental Material. We can compare these results to the error rate of ordinary two-qubit gates subject to single qubit photon losses. In these gates, in absence of control and leakage errors, increasing the gate duration always increases the error rate through proliferation of single qubit errors. In contrast, for the XCX operation between VSLQ copies, doubling the gate duration decreases the error rate once $T_{1P}$ is sufficiently high, likely due to suppression of higher order processes (single-VSLQ logical error is a nearly negligible contribution here). The resulting gate error rate for 400 ns XCX with $T_{1P} = 64 {\rm \mu s}$ is about 85\% of the error rate for an ordinary two-qubit gate of one tenth the duration. An elegant physical implementation of our two-qubit gate set could be based on tunable, flux-driven couplers such as those demonstrated in \cite{chenneill2014,mckayfilipp2016}; see the Supplemental Material for details.

\textit{Two-qubit gates: Error transparent phase gate:} The XCX gate of the previous section is not truly error-transparent, since the gate ceases to operate between a photon loss and its correction. To generate an error-transparent gate for the VSLQ, we first note that, if both VSLQ copies are in the logical state manifold, the entangling $Z_{LA} Z_{LB}$ operation (where the $X_L$ operators in XCX have been replaced with $Z_L$ terms) can be generated as:
\begin{eqnarray}\label{bareCZ}
H_{CZZ} \of{t} &=&  g \of{t} \of{ \tZ_{lA} \tZ_{lB} + \tZ_{rA} \tZ_{rB}  }; \\
&\simeq& - \frac{g \of{t}^2}{W} \tZ_{lA} \tZ_{lB} \tZ_{rA} \tZ_{rB}. \nonumber 
\end{eqnarray}
This coefficient is generated perturbatively, with a factor of 4 from combinatorics canceled by the energy cost $4W$ of transiently flipping both VSLQ copies into $W = -1$ states from a $\tZ_{lA} \tZ_{lB}$ or $\tZ_{rA} \tZ_{rB}$ term. Now imagine that a single photon is lost in one of the VSLQ copies. The action of the $ZZ$ terms now only has an energy cost of $2 W$, which suggests that if we define $\tZ_{i}'' \equiv P_{i}^2 + \frac{1}{2} P_{i}^1 - P_{i}^0$ the Hamiltonian
\begin{eqnarray}\label{fixCZ}
H_{CZZ} \of{t} &\to&  g \of{t} \of{ \tZ''_{lA} \tZ''_{lB} + \tZ''_{rA} \tZ''_{rB}  } \\
&\simeq& - \frac{g \of{t}^2}{W} \tZ'_{lA} \tZ'_{lB} \tZ'_{rA} \tZ'_{rB}. \nonumber 
\end{eqnarray}
will have the same perturbative coefficient (to second order in $g$, at least) and return the same phase evolution even if a single photon is lost, as the coefficient is cut in half when acting on a $\ket{1}$ state and a $\ket{1}$ state is only present due to decay from $\ket{2}$ states (hence the replacement of $\tZ''$ with $\tZ'$). The expression (\ref{fixCZ}) is understood to only be correct if one or zero photons have been lost (from any of the four transmons); if two photons are lost the gate will not operate as intended, but at high $T_{1P}$ this is rare and the gate error will decrease nearly quadratically in increasing $T_{1P}$.

We can benchmark the performance of these gates numerically. Using the profile shape in FIG.~\ref{gatefig}, we demonstrate super-linear scaling of gate error, with the errors rate for a CZZ gate split over two or four EC pulses (200 or 400 ns total gate time) are best fit by $p \of{T_{1P}} = 0.0057 / T_{1P} + 0.253 / T_{1P}^2 $ ($p=1.48\times10^{-4}$ at $T_{1P} = 64 {\rm \mu s}$) and $0.0064 / T_{1P} + 0.380 / T_{1P}^2 $, respectively. The quadratic term thus dominates until $T_{1P}$ is large. We attribute the linear term to higher order perturbative corrections; gate error in the absence of random error processes is of order $10^{-7}$. Gate waveforms for XCX/CZZ are simple Gaussian/Tanh pulses, and gate error could undoubtedly be further reduced in both cases through sophisticated numerical optimization.
 
\textbf{State measurement and conclusions:} To measure the state of a VSLQ, we adopt the protocol of Didier \textit{et al} \cite{didierbourassa2015}, and implement a coupling between each qubit and a common readout resonator, of the form
\begin{eqnarray}\label{HM}
H_M = m \of{t} \of{ \tX_l + \tX_r } \of{a_R^\dagger + a_R}.
\end{eqnarray}
To measure the state, we ramp $m \of{t}$ from zero to a finite value and then measure the resulting resonator signal. In absence of photon losses, this tracks the $\tX$ eigenvalue, and the phase of the resonator will evolve toward the target value. If a single photon is lost, one of the $\tX$ operators will return zero, but the other will continue to operate normally, and the pointer state will evolve in the appropriate direction (though at half the rate). One can thus accurately capture the $\tX$ eigenvalue of the parent logical state by simply measuring for a long enough time to achieve the appropriate contrast even with the drive strength cut in half. Ignoring higher order corrections (that can in general be suppressed through pulse shaping or by adding further signals) we thus expect measurement error to scale as $T_{M} T_R /T_{1P}^2$, where $T_M$ is the characteristic measurement time that depends on the resonator damping rate $\kappa$ and other experimental considerations. One can achieve similar scaling by measuring $\tX_l$ and $\tX_r$ independently, or by mapping the $\tX_{l/r}$ eigenvalues to the state of two ordinary transmons, which are then measured by dispersive shift.

We have presented a universal, error corrected quantum gate set for the VSLQ architecture, in which gate operations inherit much of the parent device's tolerance to single-qubit errors. The simulated performance of these gates is extremely promising, with two-qubit gate error rates in the low $10^{-4}$ range achievable without further increasing $T_{1}$ beyond what has already been achieved in contemporary experiments. Combined with robust measurement protocols, we have outlined the essential ingredients required for a ``dissipative subsystem code," where VSLQ copies replace single qubits in a topological code, improving the fidelity of each code operation by an order of magnitude. However, it is important to caution that leakage (short-lived populations of $\ket{1_l 1_r}$ or $\tX_l \tX_r =-1$ states) must be rigorously analyzed before making quantitative predictions about code performance.

\section{Acknowledgements}
We would like to thank Jonathan DuBois, Eric Holland, David Rodriguez-Perez, Yaniv Rosen and David Schuster for useful discussions. This work was supported by Lawrence Livermore National Laboratory and by the Louisiana Board of Regents RCS grant (LEQSF(2016-19)-RD-A-19.

\bibliography{biblio,EC_bib,SLbib}

\section{Supplemental Material}

\textit{Benchmarking gates:} To benchmark our gates, we adopted the following protocol. For a given $\Gamma_P$, gate waveform and error correction protocol we generated an initial density matrix by initializing each copy in the $\tX = +1$ ground state and evolving it under error correction until the decay rate equilibrates (eliminating short-time behavior related to residual shadow qubit populations and the choice of operator being measured). Empirically this occurs in fewer than ten error correction cycles, so we ran ten simulated error correction cycles to prepare initial density matrices. We store the resulting density matrix $\rho_0$, and then use idealized error-tolerant rotations to prepare each copy in one of the six canonical directions on the Bloch sphere. We then measure the initial fidelity using projection operators $\of{ 1 \pm O_L}/2$ (where $O_L$ is one of the error-transparent $X_L$, $Y_L$ or $Z_L$), evolve the system for one full gate operation, apply the ideal transformation (\ref{defXCX}) (or its ZZ equivalent) to invert the physical gate, and then measure the projection operators again. The difference in fidelities, averaged over all 36 initial directions, yields the average error per two-qubit gate. We chose this approach over simulating randomized benchmarking because of the large computational cost of numerically integrating the Lindblad equation, given that the Hilbert space of two VSLQ copies and their attendant shadow qubits contains $36^2 = 1296$ elements; this method requires significantly fewer Lindblad evolutions.

\begin{figure}
\includegraphics[width=3.25in]{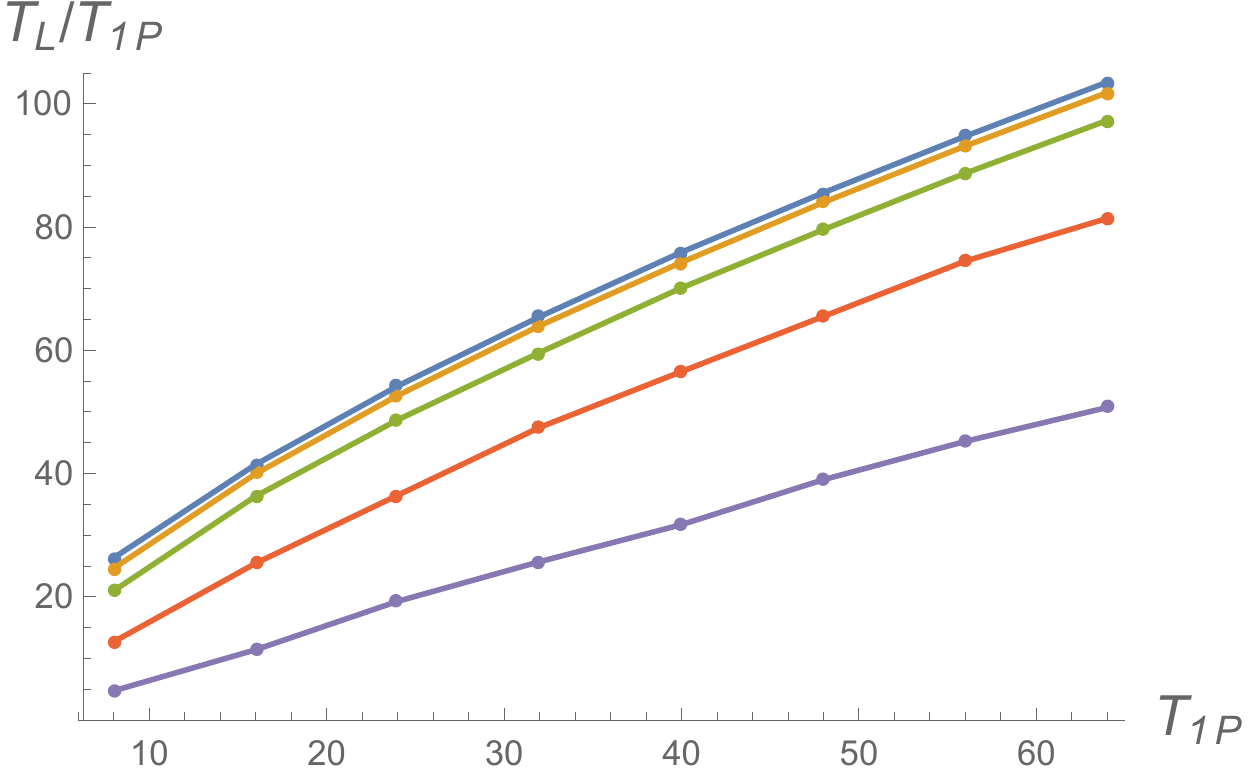}
\caption{Lifetime of $\tX$ eigenstates under photon losses and $1/f$ phase nosie. We here plot the extracted lifetime improvement, $T_L / T_{1P}$ of a $\tX$ eigenstate under photon losses with a rate $1/T_{1P}$ and $1/f$ phase noise affecting both qubits with an average strength chosen such that the single qubit Ramsey $T_{2R}$ (free induction decay, assuming no photon loss) which is infinite (blue) or $T_{2R} = \cuof{1,1/2,1/4,1/8} T_{1P}$ (top to bottom; gold, green, red and purple). Note that the transmons in the VSLQ experience twice the effective phase noise of a single qubit because the logical states are in the two-photon manifold. The lifetime is computed by numerically integrating the Lindblad equations with randomly fluctuating $h_{l/r} \of{t} a_{l/r}^\dagger a_{l/r}$ terms added, with the trajectory averaged over 400 random pairs of phase error signals per data point. Since single-VSLQ error is a small fraction of the total error in multi-qubit gates, and even relatively strong $1/f$ noise does not prevent large lifetime increases, we are justified in neglecting $1/f$ noise in our two-qubit gate simulations. The lifetime of $Y_L$ and $Z_L$ eigenstates is slightly less than half the $\tX$ lifetime as additional error channels contribute there. $\Omega$ and $\Gamma_S$ vary in the ranges $2\pi \times \cuof{2.63,1.24}$ and $\cuof{23.3,11.0}$, respectively; both decrease as $T_{1P}$ increases, reducing the contribution of errors induced by the shadow resonators themselves.}\label{1overf}
\end{figure}

\textit{Phase noise:}
We quantify the resilience of the VSLQ to $1/f$ phase noise in FIG.~\ref{1overf}. As predicted in the original work, the combination of a strong driving term (the 25 MHz $W$ coupling) and rapid error correction suppress the effect of phase noise, so that even relatively strong phase noise does not prevent dramatic increases in state lifetime. To compute these values, we initialized a single VSLQ copy in an a $\tX$ eigenstate, and allowed it to evolve under photon losses, continuous passive error correction and $1/f$ phase noise acting on each of the two primary qubits, with $\avg{\tX \of{t}}$ averaged over 400 random noise traces before fitting to extract a lifetime. We can see from these results that even a modest $T_{2R}$ of a few $\mu$s still allows for order of magnitude increases in state lifetime, and since single VSLQ logical errors are a small contribution to the total two-qubit gate error rate, we expect that the high two-qubit gate fidelities derived elsewhere in this work will be only modestly affected by phase noise. This is reassuring, since we expect that the flux loop couplers integral to the VSLQ's design will be an additional source of $1/f$ flux noise, so the bare qubit $T_{1P}$ and $T_{2R}$ (in absence of drives and dissipation) may be somewhat lower than single qubits fabricated using the same process and subjected to the same environment.

\textit{Implementing operations: $P_{l}^1 \tX_r$}

To implement $P_{l}^1 \tX_r$ we apply two tones, one through the central coupler of the VSLQ and another which is applied directly to the charge or flux degree of freedom of the right transmon. We assume the central coupler has a bias of $\pi/2$ so that all the terms which generate $W$ show up at first or third order (if the central coupler has a 0 or $\pi$ bias the signal structure will change but the result will be the same). We add to the Hamiltonian a term of the form:
\begin{eqnarray}
\of{ \Omega_{1} \cos \phi_l \sin \phi_r + \Omega_{2} \sin \phi_r} \cos \sqof{2 \pi \of{\omega - \delta / 2} t} 
\end{eqnarray}
Optionally, the $\Omega_2$ term can be implemented through coupling to $Q_r$ instead of $\sin \phi_r$. The single photon transition induced by this drive is off-resonant, but the two-photon transition produced by squaring this operator is not, so we can treat it perturbatively. The result, taking into account mixing with higher levels, takes the form:
\begin{eqnarray}
H_{eff} = g \frac{\of{ \Omega_{1} \cos \phi_l + \Omega_{2}}^2}{\delta} \of{d_1 \tX_r + d_2 \tZ_r}
\end{eqnarray}
Here, $d_1$, $d_2$ and $g$ are dimensionless prefactors that need to be computed in perturbation theory. For realistic parameters $d_1 \gg d_2$ so we can effectively ignore $d_2$ (or cancel it through the method described below). If we then write $\cos \phi_l$ as a diagonal matrix by carefully choosing the ratio of $\Omega_{1}$ to $\Omega_{2}$ we can obtain $H_{eff} = \Omega' \of{1 - P_{l}^1 } \tX_r$. By combining this with an ordinary $\tX_r$ term generated through other means (such as additional single-photon drives) we can arrive at $P_l^1 \tX_r$ as desired.

\begin{figure*}
\includegraphics[width=6.0in]{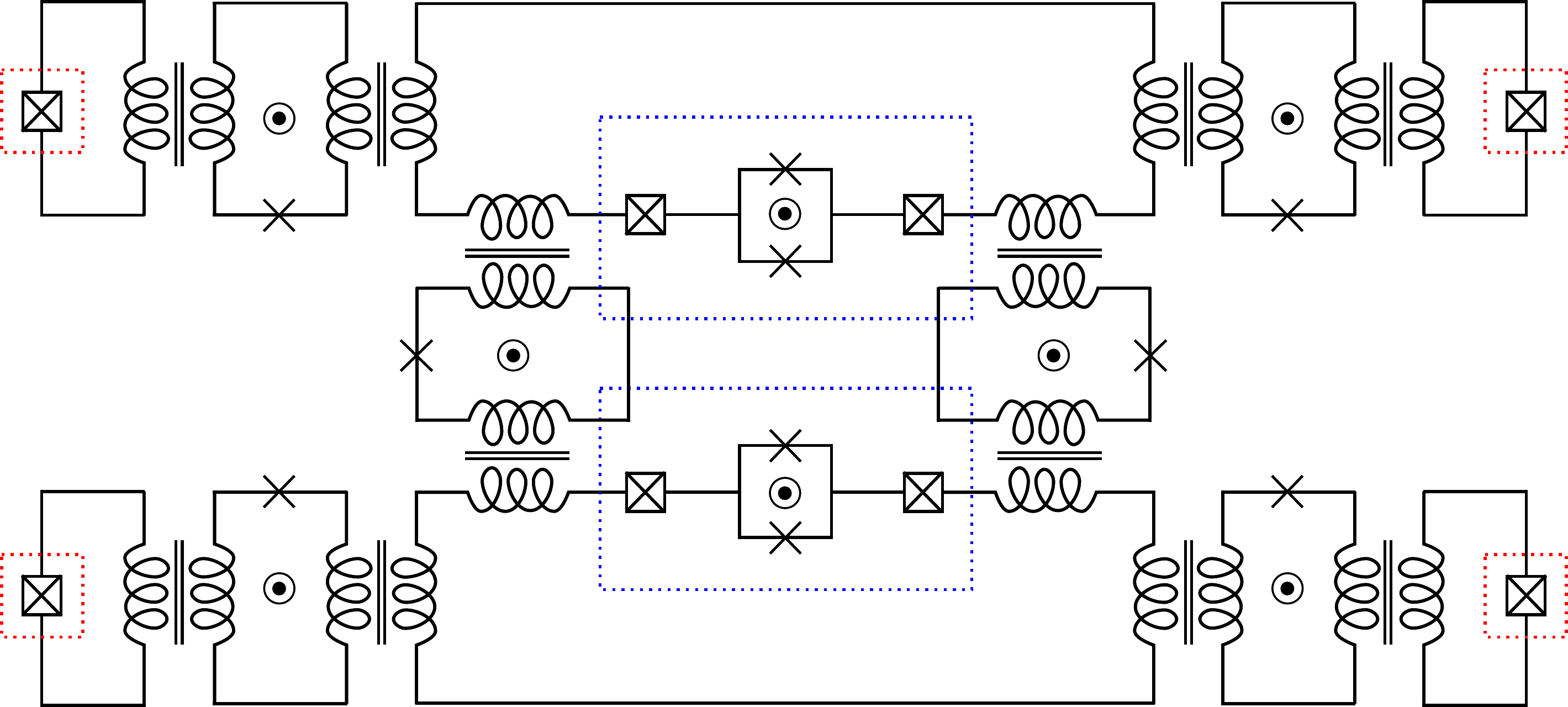}
\caption{(Color online) Example coupling structure which can implement the driven CZZ and XCX gates. The flux-driven couplers \cite{chenneill2014} are tunable mutual inductances between the primary VSLQ circuits (blue boxes) and their associated shadow qubits (red boxes). By driving these couplers at two-photon creation frequencies with appropriate detuning, the ``$++/- -$" terms required for photon loss correction and $\tZ_{lA}'' \tZ_{lB}''$ can be implemented; adding two-photon exchange terms and tuning to resonance can implement $\tX_{lA} \tX_{lB}$. Since the couplings are through mutual inductances and not Josephson junctions, flux quantization rules are straightforward and issues involving unwanted closed loops do not arise. This design could be extended to couple each VSLQ to four or more nearest neighbors (not shown), required for topological error correction codes. The VSLQ states could be measured through coupling to resonators (also not shown), either through tunable mutual inductances or a more complex lumped-element design.}
\end{figure*}

\textit{Implementing $\tZ_{lA}'' \tZ_{lB}''$, $\tZ_{l}' \tZ_{r}'$ and XCX:}

One way to implement the $\tZ_{lA}'' \tZ_{lB}''$ coupling is through a tunable mutual inductor, as in \cite{chenneill2014}, though formally any rapidly tunable coupling element could likely be adapted for our purposes. This coupling takes the form:
\begin{eqnarray}\label{HMI}
H_{AB} = f_l \of{t} \of{\phi_{lA} \phi_{lB}} + f_r \of{t} \of{\phi_{rA} \phi_{rB}}
\end{eqnarray}
Assuming the plasma frequency of the coupler junction is suitably large, these couplers can be driven at high frequencies. Further, because there is no physical current flow across the mutual inductance, no unwanted terms arise from flux quantization requirements around large loops. Such a design is thus scalable for a large system.

To achieve the $\tZ_{lA}'' \tZ_{lB}''$ (\ref{fixCZ}) Hamiltonian required for CZZ, we drive the coupler via:
\begin{eqnarray}
f_{l/r} \of{t} = \Omega_0 \cos \sqof{ 2 \pi \of{\omega_{l/rA} + \omega_{l/rB} + 2 \gamma } t}.
\end{eqnarray} 
Here, $\gamma$ is a detuning such that the drive signal is relatively far off-resonance from any two-qubit transitions. While driving the circuit in this manner does not change photon number in either qubit, off-resonant mixing with other levels creates a set of energy shifts in second order perturbation theory:
\begin{eqnarray}\label{HAB1}
H_{AB} \simeq \sum_{ij} C_{ij} P_{l/rA}^{i} P_{l/rB}^{j}
\end{eqnarray}
With a bit of algebra, we can rearrange the terms in $H_{AB}$ as
\begin{eqnarray}\label{HAB2}
H_{AB} &=& c_{1} \of{ P_{l/rA}^1 + P_{l/rB}^1} +  c_{ZZ} \tZ_{l/rA}'' \tZ_{l/rB}''   \\
& & + c_{Z} \of{ \tZ_{l/rA} + \tZ_{l/rB} } + c_{11} \of{ P_{l/rA}^1 P_{l/rB}^1} + c_0. \nonumber
\end{eqnarray}
Of these terms, the $c_{ZZ}$ coefficient is the target $g \of{t}$ in (\ref{fixCZ}), $c_0$ is a constant which does not change the system's dynamics, $c_{1}$ is an irrelevant energy shift for $\ket{1}$ states that can be compensated by adjusting the frequency of error correction drives, $c_{Z}$ is a single qubit energy shift between levels 0 and 2 that can be cancelled through other means, and $c_{11}$ is an irrelevant interaction term that only influences the system when both VSLQ copies have lost a photon, causing a gate error. Given a desired $c_{ZZ}$ and focusing on $C_{00}$, $C_{01}$, $C_{02}$, $C_{12}$ and $C_{22}$, if we equate those terms in (\ref{HAB1}) with their equivalents in (\ref{HAB2}), we have a simple system of five equations with five unknowns ($\Omega$, $\gamma$, $c_{0}$, $c_{1}$, $c_{Z}$) that we can readily solve to find $\Omega$ and $\gamma$. While the particular values solving these equations depend on the fixed device parameters $E_{J}$ and $E_{C}$ for each of the four qubits, for physically realistic $E_{J}/E_C = 75$ and $E_{J} = 2 \pi \times 18$ GHz, we obtain $c_{ZZ} = 2\pi \times 5$ MHz from $\Omega_0 \simeq 2\pi \times 75$ MHz (this is reduced to about 12 MHz when matrix elements from $\phi$ are included) and $\gamma \simeq 2\pi \times 50$ MHz. Such values are all experimentally accessible; a $c_{ZZ}$ coefficient of this strength is sufficient for the 400 ns gate in the text. 

Note that this treatment is somewhat abbreviated for simplicity and clarity, and a more sophisticated analysis would take the $W$ terms into account (off resonant driving is assumed to transiently take a state out of the $W=1$ manifold but not back into it if acting on a $\ket{1}$ state) in generating the $\tZ_{lA}" \tZ_{lB}"$ coefficients. Though we do not include it here, we have worked through such a treatment for realistic device parameters and shown that it modifies the target coefficients only slightly and does not introduce new terms which cannot be cancelled by simple single-qubit operations. We also note that the $\tZ_{l}' \tZ_{r}'$ operation required for single qubit gates can be engineered through exactly the same protocol (with the drive signal applied through the central SQUID coupler in that case), and such couplers could also be operated in different frequency regimes to enact XCX.

\end{document}